\documentclass[prb,aps,superscriptaddress,reprint
]{revtex4-2}

\usepackage[T1]{fontenc}
\usepackage{graphicx}
\usepackage{units}
\usepackage{amsmath,amssymb}
\usepackage{color}
\usepackage{upgreek}
\usepackage[colorlinks=true,urlcolor=blue,linkcolor=blue,citecolor=blue]{hyperref}
\usepackage{xcolor}

\usepackage{ulem}

\begin{document}

\title{Probing proximity-induced superconductivity in bilayer \\ graphene using gate-defined quantum dots}

\author{Philipp Schmidt}
\email[]{philipp.schmidt3@rwth-aachen.de}
\affiliation{JARA-FIT and 2nd Institute of Physics, RWTH Aachen University, 52074 Aachen, Germany}
\affiliation{Peter Gr\"unberg Institute (PGI-9), Forschungszentrum J\"ulich, 52425 J\"ulich, Germany}
\author{Mathies Funke}
\affiliation{JARA-FIT and 2nd Institute of Physics, RWTH Aachen University, 52074 Aachen, Germany}
\author{Till Tenger}
\affiliation{JARA-FIT and 2nd Institute of Physics, RWTH Aachen University, 52074 Aachen, Germany}
\author{Katarina Stanojevi\'{c}}
\affiliation{JARA-FIT and 2nd Institute of Physics, RWTH Aachen University, 52074 Aachen, Germany}
\author{Hubert Dulisch}
\affiliation{JARA-FIT and 2nd Institute of Physics, RWTH Aachen University, 52074 Aachen, Germany}
\affiliation{Peter Gr\"unberg Institute (PGI-9), Forschungszentrum J\"ulich, 52425 J\"ulich, Germany}
\author{Kenji Watanabe}
\affiliation{Research Center for Electronic and Optical Materials, National Institute for Materials Science, 1-1 Namiki, Tsukuba 305-0044, Japan}
\author{Takashi Taniguchi}
\affiliation{Research Center for Materials Nanoarchitectonics, National Institute for Materials Science,  1-1 Namiki, Tsukuba 305-0044, Japan}
\author{Fabian Hassler}
\affiliation{Institute for Quantum Information, RWTH Aachen University, 52056 Aachen, Germany}
\author{Christian Volk}
\affiliation{JARA-FIT and 2nd Institute of Physics, RWTH Aachen University, 52074 Aachen, Germany}
\affiliation{Peter Gr\"unberg Institute (PGI-9), Forschungszentrum J\"ulich, 52425 J\"ulich, Germany}
\author{Szabolcs Csonka}
\affiliation{Department of Physics, Institute of Physics, Budapest University of Technology and Economics, M\"uegyetem rkp. 3., H-1111 Budapest, Hungary}
\affiliation{MTA-BME Superconducting Nanoelectronics Momentum Research Group, M\"uegyetem rkp. 3., H-1111 Budapest, Hungary}
\author{Christoph Stampfer}
\affiliation{JARA-FIT and 2nd Institute of Physics, RWTH Aachen University, 52074 Aachen, Germany}
\affiliation{Peter Gr\"unberg Institute (PGI-9), Forschungszentrum J\"ulich, 52425 J\"ulich, Germany}

\date{\today}

\begin{abstract}

Van der Waals heterostructures enable the combination of two-dimensional (2D) materials with different electronic properties in a single device. Bilayer graphene (BLG) is particularly attractive as its electrically tunable band gap enables local control of tunnel barriers and quantum dots (QDs). Here, we realize an all-2D hybrid platform based on BLG proximitized by superconducting NbSe$_2$. Using electrostatic gates, we define tunnel barriers and QDs at different distances from the superconductor-semiconductor interface. The QDs serve as local spectroscopic probes of the proximitized BLG channel, forming tunable superconductor--quantum dot--normal conductor junctions. Coulomb blockade and finite-bias spectroscopy reveal a proximity-induced superconducting gap of up to $80\,\mathrm{\mu eV}$ and allow to track its evolution with distance from the NbSe$_2$ contact. The local density of states remains suppressed over distances exceeding 1\,$\mu$m.
Our results demonstrate the potential of BLG--superconductor hybrids as a tunable platform combining QDs and quantum point contacts with superconductivity.
\end{abstract}


\maketitle

Gate-tunable superconductor-semiconductor hybrids
combine superconducting correlations with electrostatically defined barriers, quantum dots and mesoscopic transport channels.
When superconductors are coupled to low-disorder semiconductors, superconducting correlations can extend over distances of several hundred nanometers via the proximity effect~\cite{wang2024oct, junger2019jul}. In particular, ballistic semiconductor nanowires contacted by superconductors exhibit a robust proximity effect and host a rich spectrum of subgap excitations~\cite{defranceschi2010oct, higginbotham2015dec, vandriel2024apr, poschl2022oct, levajac2023oct, jellinggaard2016aug, steffensen2022apr, grove-rasmussen2018jun, estradasaldana2020jul, su2017sep, sherman2017mar, kurtossy2021oct, scherubl2020apr, junger2023jul}, including zero-bias conductance peaks~\cite{mourik2012may, das2012dec, deng2016dec, gul2018mar, chen2019sep, vaitiekenas2021jan}. 
These systems have enabled the realization of proximitized quantum dots, which provide access to functionalities such as Cooper pair splitting~\cite{wang2022dec, wang2023aug, kurtossy2022sep, scherubl2022may, hofstetter2009oct, baba2018jun, bordoloi2022dec}, and can serve as building blocks for engineered topological superconducting phases, including Kitaev-chain–like systems~\cite{dvir2023feb, zatelli2024sep, tenhaaf2024jun}.
More broadly, these advances open new avenues for exploring unconventional superconducting states and offer promising routes towards quantum coherent devices, including potential qubit implementations~\cite{kitaev2003jan, chtchelkatchev2003jun, wang2019feb}.

Van der Waals heterostructures based on two-dimensional (2D) materials extend this platform by enabling the assembly of superconductor–semiconductor hybrids with atomically sharp and in principal contamination-free interfaces. In this context, graphene and bilayer graphene proximitized by the layered superconductor NbSe$_2$ have emerged as
attractive systems, combining induced superconductivity with the electrostatic tunability of graphene~\cite{sahu2018aug, efetov2016apr, dvir2021mar, devidas2023mar, zalic2023jul, han2018feb, li2020may}. This combination enables access to regimes that are difficult to realize in conventional semiconductor platforms, with continuous control over carrier density, band structure, and coupling to the superconducting reservoir, while providing access to both the conduction and valence bands within the same device.
Bilayer graphene (BLG) has recently emerged as a highly controllable system for gate-defined quantum point contacts \cite{overweg2018jan, Kraft2018Dec, banszerus2020may} and quantum dots \cite{eich2018jul, banszerus2018aug}, with potential for hosting spin and valley qubits. 
Furthermore, electron and hole quantum dots in BLG can be defined and controlled purely by electrostatic gating \cite{banszerus2023jun, banszerus2020oct, banszerus2018aug}. This enables the realization of complex hybrid mesoscopic structures within a single material system and provides a route towards engineered superconducting and potentially topological quantum devices~\cite{leijnse2012oct}.

\begin{figure*}
    \centering
    \includegraphics[width=0.95\linewidth]{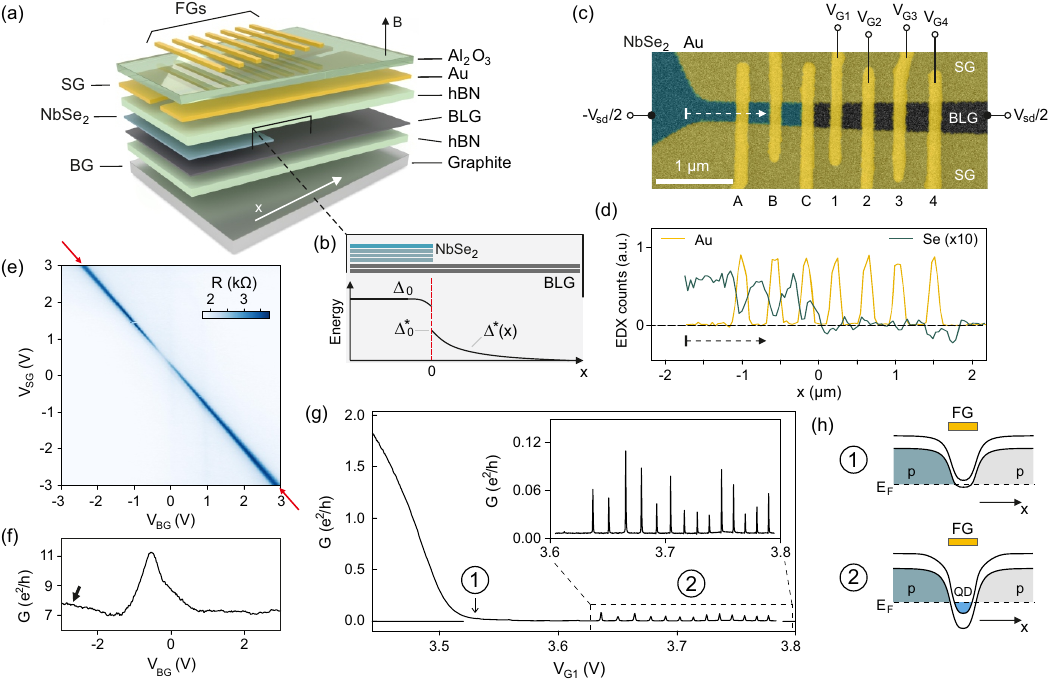}
    \caption{
    (a) Schematic three-dimensional cross-section of the device structure. The split gates (SGs) define the conducting channel, while voltages applied to the finger gates (FGs) allow to locally control the conductance through the channel.
    (b) Cross-section of the layer stack along the $x$-direction, highlighting the decay of the proximity induced gap $\Delta^*$.
    (c) False colored scanning electron microscopy image showing the gate structures of the device.
    (d) Normalized energy-dispersive X-ray spectroscopy (Au-M and Se-L series) along the dotted white line in panel (c) indicating the position of the NbSe$_2$ flake with respect to the Au FGs.
    (e) Resistance $R$ as function of back gate voltage $V_\mathrm{BG}$ and split gate voltage $V_\mathrm{SG}$ measured at a small out-of-plane magnetic field $B=20\,\mathrm{mT}$.
    (f) Line-cut along the diagonal feature of increased resistance, i.e. reduced conductance (indicated by the red arrow in panel (e)).
    (g) Conductance as function of FG~1 voltage, $V_{\mathrm{G1}}$, measured at $V_\mathrm{BG}=-2.4\,\mathrm{V}$ and $V_\mathrm{SG}=2.9\,\mathrm{V}$, highlighting both, the tunneling (1) and quantum dot regime (2). Inset: Coulomb resonances in the quantum dot regime.
    (h) Schematics of the valence and conduction band edge profiles along the p-doped channel, illustrating the tunneling regime (1) and the quantum dot (QD) regime (2), tuned by the FG voltage.
    }
    \label{fig:fig1}
\end{figure*}

\begin{figure*}
    \centering
    \includegraphics[width=0.95\linewidth]{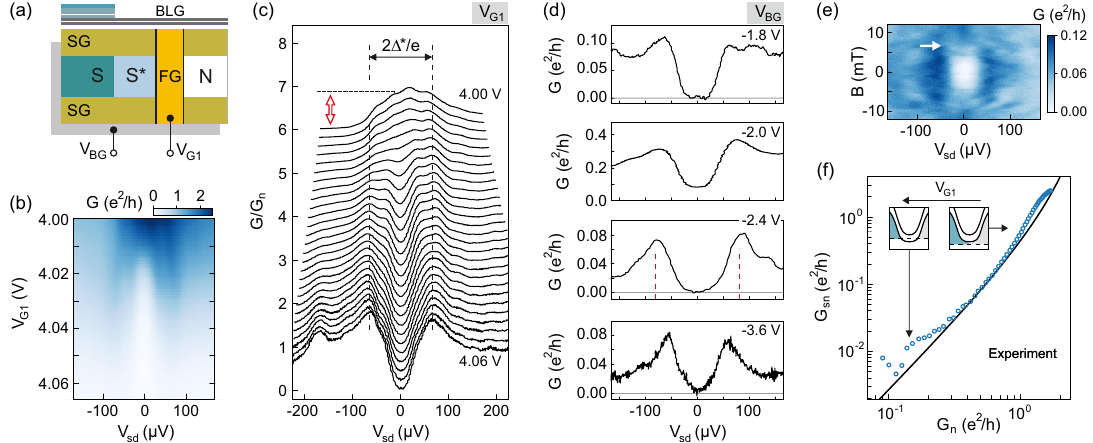}
    \caption{
        (a)~Schematic cross section of the NbSe$_2$/BLG heterostructure (top) and top-view device layout showing the gate geometry, superconducting contact S, proximitized BLG region S$^*$, and normal BLG region N (bottom).
        (b)~Differential conductance $G=dI/dV_{\mathrm{sd}}=I_{\mathrm{AC}}/V_{\mathrm{AC}}$ as a function of source-drain voltage and FG voltage, showing the transition from the open-channel to the tunneling regime. Data are measured at $V_\mathrm{BG}=-2.75~\mathrm{V}$ and $V_\mathrm{SG}=3.35~\mathrm{V}$.
        (c)~Normalized differential-conductance traces (see text for details) at selected finger-gate voltages, showing suppressed conductance for $|V_\mathrm{sd}|<\Delta^*/e$ in the tunneling regime and enhanced conductance in the open-channel regime.
        (d)~Conductance as as a function of $V_{\mathrm{sd}}$ in the tunneling regime measured at different back-gate voltages (see labels).
        (e)~Differential conductance in the tunneling regime ($V_\mathrm{G1}=4.06~\mathrm{V}$) as a function of $V_{\mathrm{sd}}$ and out-of-plane magnetic field.
        (f)~Zero-bias differential conductance, $G_{\mathrm{sn}} = G(V_\mathrm{sd}=0)$, plotted against the normal state conductance, $G_{\mathrm{sn}} = G(V_\mathrm{sd}=150~\mathrm{\mu V})$ for the data in panel (b). The dashed line indicates the theoretical prediction. Insets show simplified band-edge profiles for the open-channel and tunneling regimes.
    }
    \label{fig:fig2}
\end{figure*}

Here, we present a versatile all-2D superconductor–semiconductor hybrid platform based on NbSe$_2$ and gate-tunable bilayer graphene, in which tunneling barriers and quantum dots (QDs) can be 
defined electrostatically by local gates positioned at different distances from the lateral superconductor–semiconductor interface.
In particular, we use the QDs as spectroscopic probes to study the density of states in the proximitized bilayer graphene reservoir (S*–region), thereby realizing a tunable superconductor–quantum dot–normal conductor (S–S*–QD–N) system with different separations between the superconductor and the quantum dot.
This approach enables direct measurements of the position-dependent proximity-induced superconducting gap in the BLG/NbSe$_2$ heterostructure. We observe a finite induced gap persisting over distances of up to $1\,\mathrm{\mu m}$.

The device structure is illustrated in Fig.~\ref{fig:fig1}(a). It consists of partially overlapping NbSe$_2$ and BLG flakes (see also top of  Fig.~\ref{fig:fig1}(b)) encapsulated in hexagonal boron nitride (hBN), assembled using a dry transfer technique and placed on a graphite back gate (BG).
Ohmic contacts to the BLG were defined by reactive ion etching followed by Cr/Au metallization, while electrical contacts to the NbSe$_2$ flake were established via patterned graphite leads, enabling a quasi-four-terminal measurement configuration (see Supporting Information, Section S1).
Split gates (SGs) and finger gates (FGs) were fabricated on the hBN/NbSe$_2$/BLG heterostructure by electron-beam lithography and subsequent Cr/Au deposition and electrically isolated by an atomic-layer-deposited Al$_2$O$_3$ dielectric.
The SGs define a conducting channel with a width in the range of $\approx$~290-370~nm, while the FGs have a width of $\approx$~150~nm and are spaced by $\approx$~420~nm.
A false-color scanning electron microscopy image of the device is shown in Fig.~\ref{fig:fig1}(c).
The approximate position of the edge of the NbSe$_2$ flake on the BLG (see red dashed line in Fig.~\ref{fig:fig1}(b)), which approximately coincides with the right edge of gate C, was identified using energy-dispersive X-ray spectroscopy (EDX), as shown in Fig.~\ref{fig:fig1}(d).
The characterization of the superconducting NbSe$_2$ flake is presented in Supporting Information, Section S3.
All transport measurements were performed in a dilution refrigerator at a base temperature of approximately 80~mK, using both DC and low-frequency lock-in techniques.

Figure~\ref{fig:fig1}(e) shows the resistance ($R$) as a function of the back gate ($V_{\mathrm{BG}}$) and split gate ($V_{\mathrm{SG}}$) voltages, with all finger gates grounded. Applying a transverse displacement field via $V_{\mathrm{SG}}$ and $V_{\mathrm{BG}}$ of opposite polarity induces a band gap in the BLG regions beneath the SG not covered by NbSe$_2$, manifested as a diagonal line of increased resistance.
In agreement with earlier work~\cite{banszerus2018aug}, this results in charge carrier confinement in the channel defined by the two split gates, 
while the carrier density within the channel is controlled by the back gate. This is reflected in the increasing conductance with increasing $V_{\mathrm{BG}}$, while the electrochemical potential in the regions beneath the split gates remains within the band gap (see arrow in Fig.~\ref{fig:fig1}(f)).
The gate lever arm is extracted from a Landau fan diagram measurement, as described in Supporting Information, Section~S2.

We next fix the back gate and split gate voltages to form a p-doped channel (see arrow in Fig.~\ref{fig:fig1}(f)), thereby matching the p-type doping induced at the NbSe$_2$ interface~\cite{li2020may}.
In this configuration, FGs 1–4 (see labels in Fig.~\ref{fig:fig1}(c)) can be used to locally tune the carrier density and, consequently, the conductance through the channel, as illustrated for FG 1 in Fig.~\ref{fig:fig1}(g). In contrast, FGs A–C are effectively screened by the $\approx 10\,$nm thick NbSe$_2$ flake and do not provide measurable electrostatic control, consistent with the absence of any observable tuning in the experiment.
We now turn to the measurement shown in Fig.~\ref{fig:fig1}(g), where the conductance is plotted as a function of the voltage applied to FG~1, $V_{\mathrm{G1}}$, at $B = 20\,\mathrm{mT}$. 
As $V_{\mathrm{G1}}$ is increased, the Fermi level is tuned across the valence band edge, leading to the formation of a tunneling barrier and a pronounced suppression of the conductance (configuration (1); see schematic illustration in Fig.~\ref{fig:fig1}(h)). As the gate voltage is increased further, an electron QD forms that is tunnel-coupled to the leads (configuration (2); see dashed box in Fig.~\ref{fig:fig1}(g) and corresponding schematic illustration in Fig.~\ref{fig:fig1}(h)). This results in a strong suppression of the conductance accompanied by distinct Coulomb peaks (see inset of Fig.~\ref{fig:fig1}(g)), which are a hallmark of sequential tunneling through a QD.

\begin{figure*}
    \centering
    \includegraphics[width=0.95\linewidth]{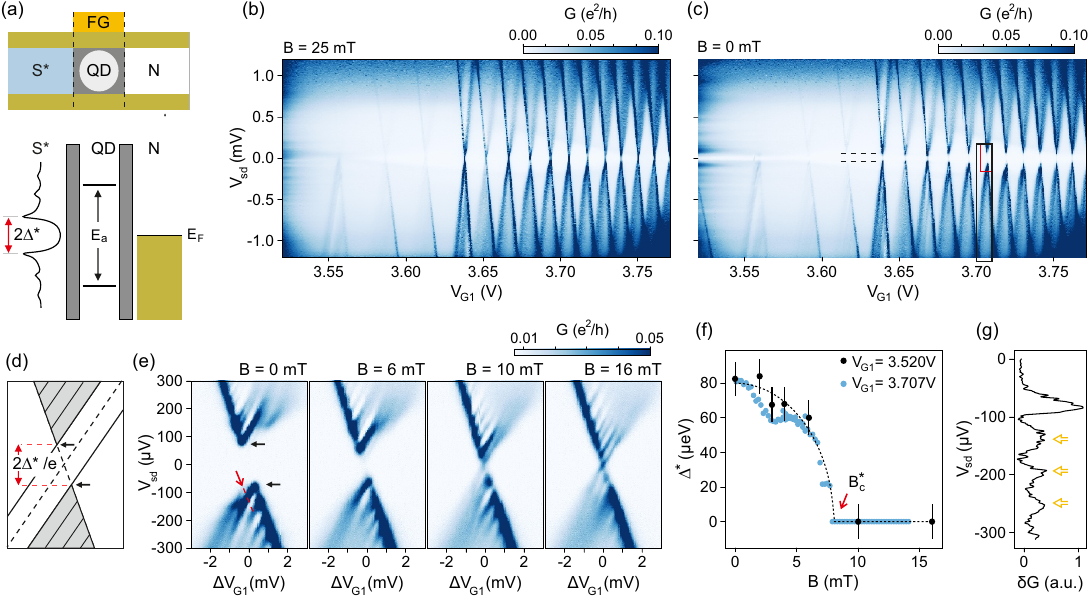}
    \caption{
        (a) Top: Schematic of the device geometry highlighting the QD formed beneath FG~1 and its coupling to the proximitized BLG region (S$^*$) on the left and the normal BLG lead (N) on the right. Bottom: Corresponding energy diagram showing the density of states in S$^*$, with $\Delta^*$ denoting the proximity-induced gap, the gate-tunable QD levels separated by the addition energy $E_\mathrm{a}$, and the Fermi level of the normal BLG lead.
        (b)~Coulomb diamonds of the few-electron QD formed beneath FG~1, measured at $V_\mathrm{BG}=-2.4~\mathrm{V}$, $V_\mathrm{SG}=2.9~\mathrm{V}$, and $B=25~\mathrm{mT}$, where proximity-induced superconductivity is suppressed.
        (c)~Same measurement as in panel (b), but at $B=0~\mathrm{mT}$, revealing a pronounced suppression of the tunneling current around $V_\mathrm{sd}=0$ due to the proximity-induced superconducting gap.
        (d) Schematic of the expected conductance (gray region) as a function of source-drain bias $V_\mathrm{sd}$ and gate voltage, illustrating the effect of the proximity-induced gap $\Delta^*$.
        (e)~Close-up of the low-bias region of the Coulomb diamond shown in panel (c), corresponding to the orange box, for varying magnetic field.
        (f)~Extracted proximity-induced gap $\Delta^*$ as a function of perpendicular magnetic field $B$ extracted from similar data as shown in panel (e) [black dots] and measurements taken at $V_{\mathrm{G1}}=3.52$~V (see panel (c)) [blue dots]. The dashed line represents a BCS-like suppression of $\Delta^*$ with a critical field $B_\mathrm{c}^*=8\,\mathrm{mT}$ (see text for details).
        (g)~Conductance traces taken along the direction indicated by the red arrow in panel (e).
    }
    \label{fig:fig3}
\end{figure*}

We first focus on configuration (1), corresponding to the tunneling regime, to investigate the superconducting proximity effect in the S$^*$ region (see Fig.~\ref{fig:fig2}(a)).
In Fig.~\ref{fig:fig2}(b) we show the differential conductance $G=dI/dV_{\text{sd}}$ as function of $V_{\mathrm{G1}}$ and $V_{\mathrm{sd}}$.
Additional tunneling spectroscopy data are shown in Supporting Information Section S4.
By increasing the finger gate voltage $V_{\mathrm{G1}}$ we smoothly transition the system from an open channel regime at $V_{\mathrm{G1}} = 4.00\,\mathrm{V}$ to the tunneling regime ($V_{\mathrm{G1}} \approx 4.06\,\mathrm{V}$), where $G$ is strongly suppressed for $|V_{\mathrm{sd}}| < 60\,\mu\mathrm{V}$.
The corresponding normalized differential-conductance traces, $G/G_\mathrm{n}$, are shown in Fig.~\ref{fig:fig2}(c), where $G_\mathrm{n}=G(V_\mathrm{sd}=150\,\mu\mathrm{V})$ is taken as the normal-state differential conductance. 
In the open regime, a conductance enhancement is observed at low bias (see red arrows), while in the tunneling regime the conductance is strongly suppressed for $|V_{\mathrm{sd}}| < 60\,\mu\mathrm{V}$. 
From these measurements, we estimate a proximity-induced gap in the density of states of BLG at the position of FG 1 of $\Delta^* \approx 60\,\mu\mathrm{eV}$.
In Fig.~\ref{fig:fig2}(d) we show tunneling traces measured at varying back gate voltage (re-tuning $V_\mathrm{SG}$ and $V_\mathrm{G1}$ to the tunneling regime) and thus varying the charge carrier density in the channel. 
We find a variation of the extracted proximity gap of $\approx 30\, \%$ across the investigated gate voltage range and the largest gap of $\Delta^* \approx 80\,\mathrm{\mu eV}$ for $V_\mathrm{BG}=-2.4,\mathrm{V}$.
We note that the proximity-induced superconducting gap is approximately an order of magnitude smaller than the parent NbSe$_2$ gap of $\Delta\approx 1$~meV~\cite{dvir2018feb,devidas2021aug}. 
Similarly reduced gaps of $\Delta^*\approx 50$--$60~\mu\mathrm{eV}$ and $240$--$360~\mu\mathrm{eV}$ have also been reported in experiments on graphene--NbSe$_2$ heterostructures~\cite{moriya2020feb,li2020may}. 
Possible origins include momentum-space mismatch between graphene and NbSe$_2$, which is predicted to depend strongly on the relative twist angle~\cite{gani2019jun}, as well as interfacial disorder and variations in interlayer separation. 
The relative contributions of these mechanisms, however, remain unclear.
We further show measurements of the differential conductance as a function of bias voltage and an external out-of-plane magnetic field $B$ (see Fig.~\ref{fig:fig2}(e)) to investigate the robustness of the induced gap in a magnetic field. 
The characteristic gap feature is suppressed at an out-of-plane magnetic field of \(B^\ast_{\rm c}\simeq 8\,\mathrm{mT}\), see Fig.~\ref{fig:fig2}(e). This field scale, which is much smaller than the critical field of the parent NbSe\(_2\) superconductor ($\approx4\,\mathrm{T}$), is expected to be set by orbital dephasing of the proximitized BLG segment, i.e. the S$^*$ region in Fig.~\ref{fig:fig2}(a). Estimating the relevant Andreev-cavity area from \(B_\mathrm{c} ^\ast A_{\rm eff}\sim \Phi_0=h/2e\) gives \(A_{\rm eff}\simeq 0.26\,\mu\mathrm{m}^2\), corresponding to an effective length \(\ell_{\rm eff}\simeq \sqrt{A_{\rm eff}}\simeq 0.51\,\mu\mathrm{m}\). This is only moderately larger than the nominal geometric length of the proximitized channel segment, \(L\simeq 300\,\mathrm{nm}\). Considering that the relevant area is set by the actual electron--hole trajectories rather than by the lithographic dimensions alone, this agreement is reasonable and supports an orbital-dephasing interpretation of the gap suppression.

In Figure~\ref{fig:fig2}(f), we plot the zero-bias conductance $G_\mathrm{sn}=G(V_\mathrm{sd}=0)$ in the superconducting state as a function of the normal-state conductance $G_\mathrm{n}$.
Notably, the data are described reasonably well by a model assuming perfect Andreev reflection at the S$^*$–N interface and single-mode transport through the BLG channel.
In this case the conductance, when the S$^*$ region  is superconducting, denoted by $G_\mathrm{sn}$ is given by 
$G_\mathrm{sn} = 2G_0 {G_\mathrm{n}^2}/{(2G_0-G_\mathrm{n})^2}$, where $G_0=2e^2/h$ and $G_\mathrm{n}$ is the conductance in the normal state~\cite{beenakker1992nov,kjaergaard2016sep}. 
The black line in Fig.~\ref{fig:fig2}(f) shows this prediction of $G_\mathrm{sn}$ as function of $G_\mathrm{n}$ with no free parameters. 
The reasonable agreement between the model and the experimental data indicates that the observed conductance evolution is consistent with the expected channel-transmission dependence of Andreev reflection. 
At low conductance, the excess $G_\mathrm{sn}$ may reflect a residual normal-transport contribution, for example from a finite subgap density of states, leakage, or parallel transport channels. 
A similar systematic excess conductance was reported in InGaAs/InAs semiconductor--superconductor devices and attributed to a small remanent normal-scattering probability~\cite{kjaergaard2016sep}. 
At high conductance, additional transport channels or mode mixing may account for the departure from the ideal single-channel description.

We next turn to configuration (2), in which an electron quantum dot is formed beneath the finger gate in the BLG channel and tunnel-coupled to the leads (see lower schematic in Fig.~\ref{fig:fig1}(h) and Fig.~\ref{fig:fig3}(a)).
We access the relevant energy scales depicted in Fig.~\ref{fig:fig3}(a) by bias spectroscopy of the gate-defined quantum dot, comparing the response for a normal-conducting left lead  at $ B = 25\,\mathrm{mT}$ with that for a superconducting left lead ($\mathrm{S}^*$) at $ B = 0$.
In the normal-state regime, the QD device exhibits well-defined Coulomb diamonds (Fig.~\ref{fig:fig3}(b)), with addition energies of approximately $E_\mathrm{a}\approx  1\,\mathrm{meV}$ (dominated by the charging energy) at low electron occupation that gradually decrease with increasing filling, in line with previous reports on gate-defined BLG QDs~\cite{eich2018jul, banszerus2018aug,banszerus2020oct}.
When the left lead is in the proximity-induced superconducting state, the overall Coulomb-blockade diamond structure remains clearly visible but is accompanied by a pronounced gap around $V_\mathrm{sd}=0$ (Fig.~\ref{fig:fig3}(c)). This behavior is consistent with sequential tunneling through a weakly coupled S$^*$--QD--N configuration (Fig.~\ref{fig:fig3}(a)) \cite{junger2019jul}. Additional Coulomb diamond spectroscopy data are shown in Supporting Information Section S6.

Before analyzing the experimental data in more detail, we first discuss, with the help of a schematic (Fig.~\ref{fig:fig3}(d)), how the proximity-induced superconducting gap $\Delta^*$ modifies the Coulomb-diamond pattern. The gap suppresses quasiparticle transport around $V_\mathrm{sd}=0$, shifting the onset of tunneling to finite source-drain bias and thereby modifying the boundaries of the Coulomb diamonds.
As a result, the tips of the Coulomb-blockade diamonds are displaced in source-drain voltage by $2\Delta^*/e$.
In addition, the capacitive coupling between the lead regions and the QD gives rise to a characteristic shift of the diamond tips along the finger gate voltage axis, $V_{\mathrm{G1}}$,  given by
$\Delta V_{\mathrm{G1}} = 2\Delta^*/(\alpha e),$
where $\alpha = 0.2$ denotes the slope of the Coulomb diamond.
This behavior is in good agreement with the experimental data shown in Fig.~\ref{fig:fig3}(e). The close-up reveals a source-drain bias shift of the Coulomb-diamond tips corresponding to a proximity-induced gap of approximately $80\,\mathrm{\mu eV}$, consistent with the value extracted independently from tunneling spectroscopy (see Fig.~\ref{fig:fig2}).
Upon increasing the out-of-plane magnetic field, the gap gradually closes until it is no longer resolvable. The extracted proximity-induced gap as a function of magnetic field is summarized in Fig.~\ref{fig:fig3}(f).
Its dependence can be reasonably well described by a BCS-like suppression of the proximity-induced gap, expressed by
$\Delta^\ast(B) = \Delta^\ast(B=0) \sqrt{1 - \left(B/ B_\mathrm{c}^\ast\right)^2}$ 
with a critical field of $B_\mathrm{c}\approx 8,\mathrm{mT}$ (see dashed line in Fig.~\ref{fig:fig3}(f)), also in agreement with the tunneling spectroscopy results as shown in Fig.~\ref{fig:fig2}(e).

\begin{figure}
    \centering
    \includegraphics[width=0.99\linewidth]{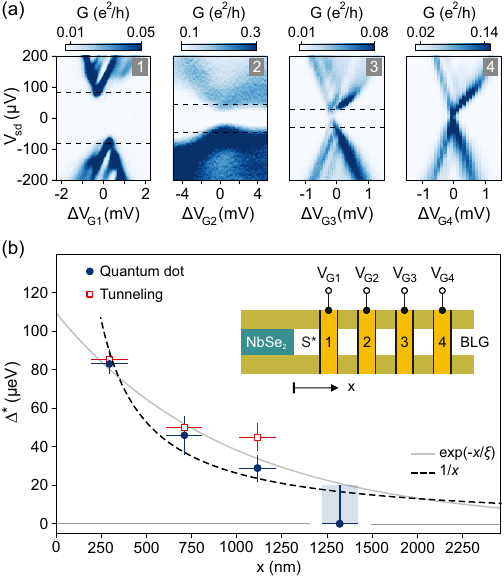}
    \caption{
    (a) Differential conductance as function of source-drain voltage $V_\mathrm{sd}$ and finger gate voltage $\Delta V_\mathrm{G{\it i}}$ ($i=1-4$) with respect to the center of the Coulomb diamond for the four different finger gates $1-4$, respectively (see labels 1 to 4 in the sub-panels and inset in panel (b)).
    (b) Proximity-induced gap~$\Delta^*$ as function of $x$, the distance from the edge of the NbSe$_2$ flake, extracted (i) from the Coulomb diamond measurements (dashed lines in panel (a)) and (ii) from tunneling spectroscopy measurements.
    The dashed black line shows the dwell-time-limited proximity gap, $\Delta^*(x)=\pi \hbar v/(4x)$ with $v\approx 5.2\times10^4\,\mathrm{m/s}$ (see text for details ).
    }
    \label{fig:fig4}
\end{figure}

In addition to the closing of the gap \(\Delta^\ast(B)\), Fig.~\ref{fig:fig3}(e) reveals a set of finite-bias resonances outside the suppressed low-bias region. 
The resonances run approximately parallel to the Coulomb-diamond edge associated with an alignment of the QD level and the electrochemical potential of the proximitized \(S^\ast\) lead region, for both bias polarities.
These resonances originate from a discrete spectrum of the finite proximitized BLG lead segment probed by the dot, rather than from ordinary QD excited states~\cite{Escott2010Jun}.
The observation of well-defined resonances is consistent with coherent Andreev states arising from finite-size quantization in the proximitized BLG segment.
For an order-of-magnitude estimate, we therefore describe
the proximitized left lead segment as a confined region with area $A$ and an orbital level spacing estimated by $\delta E \sim 4/(D(E)\,A)$, with $D(E)$ being the density of states and the 4 accounting for spin and valley degeneracy.
From a representative line cut at \(B=0\), Fig.~\ref{fig:fig3}(g), we extract a spacing of \(\Delta V_{\rm sd}\approx 55\,\mu\mathrm{V}\) (see arrows in Fig.~\ref{fig:fig3}(g)), corresponding to an energy scale of \(\delta E \approx 55\,\mu\mathrm{eV}\).
Taking this value and $D(E) \approx 0.8$~(eV nm$^2$)$^{-1}$ (see Supporting Information, Section S8) we obtain a characteristic linear dimension of $\sqrt{A} \approx 300$~nm, in agreement with the lithographic dimension of the device.
This interpretation is further supported by the observation of (i) that the energy excitations are not coupled to the proximity induced gap $\Delta^*$, meaning that they do not vanish with vanishing $\Delta^*$ (see e.g. 10~mT data shown in Fig.~\ref{fig:fig3}(e)) and (ii) that the number of visible excitations increases with small out-of-plane $B$-field, 
which is consistent with a possible lifting of the valley degeneracy associated with a sizable valley $g$-factor (Supporting Information, Section~S8), in line with earlier work~\cite{moller2023sep}.

Finally, the electrostatic tunability of the BLG platform allows us also to define QDs using different finger gates, thereby varying the distance between the QD and the lateral NbSe$_2$/BLG interface.
We use this to probe how the proximity-induced superconducting gap evolves spatially by performing bias spectroscopy measurements on quantum dots defined by the different FGs 1-4 with increasing distance of up to around $1.5\,\mathrm{\mu m}$ from the lateral superconductor/BLG interface.
For all measurements, the back- and split-gate voltages were fixed at $V_{\mathrm{BG}}=-2.4~\mathrm{V}$ and $V_{\mathrm{SG}}=2.9~\mathrm{V}$. Only the finger gate defining the respective QD was varied, while all other finger gates were held at $0~\mathrm{V}$. Each QD was tuned into a comparable low-carrier regime before performing the spectroscopy.
As shown in Fig.~\ref{fig:fig4}(a), close-ups of the corresponding Coulomb diamonds reveal a reduction of the induced gap from FG~1 to FG~3, while for FG~4 no resolvable gap feature remains. The extracted position-dependent gap $\Delta^*(x)$, including complementary values obtained from tunneling spectroscopy (see Supporting Information, Section S4), is summarized in Fig.~\ref{fig:fig4}(b).
The horizontal error bars indicate the uncertainty in the position $x$, arising from the finite width of the finger gates, while the vertical error bars reflect the uncertainty in the extraction of $\Delta^\ast$ (see Supporting Information, Section S5).

The position dependence of $\Delta^*$ roughly follows the energy scale expected for a ballistic, dwell-time-limited proximity effect. In this picture, the relevant excitation energy is set by the time of flight from the lateral NbSe$_2$/BLG interface and scales as $\Delta^*(x)\sim\pi\hbar v/(4x)$, where $x$ denotes the distance from the tunneling-barrier to the superconducting interface (where the Andreev reflection takes place) and $v$ is the effective longitudinal velocity~\cite{melsen1996jul,cserti2002aug}. This one-parameter model describes the data using $v\simeq5.2\times10^4$~m/s [dashed black line in Fig.~\ref{fig:fig4}(b)], compatible with the reduced band velocity of gapped BLG close to the band edge.
For comparison, we also consider a phenomenological exponential decay, $\Delta^*(x)=\Delta_0^*\exp(-x/\xi)$, yielding $\Delta_0^*\approx110$~$\mu$eV and $\xi\approx740$~nm [solid gray line in Fig.~\ref{fig:fig4}(b)]~\cite{natterer2016jan,ganguli2022mar}. 
While both descriptions capture the observed trend, the physically reasonable velocity obtained from the ballistic model, together with a quantum mean free path of approximately $250$~nm inferred from Shubnikov--de Haas oscillations, likely providing a conservative estimate of the transport mean free path (see Supporting Information, Section S7), favors a ballistic or quasi-ballistic interpretation, consistent with previous studies of high-quality BLG devices~\cite{banszerus2020may}. Nevertheless, diffusive contributions cannot be excluded.

While finalizing this manuscript, we became aware of related work reporting proximity-induced superconductivity in a BLG quantum point contact~\cite{Galante-Agero2026Aug, horgan2026oct}, providing complementary evidence for superconducting proximity effects in gate-defined BLG devices

In conclusion, we have demonstrated a gate-tunable BLG platform coupled to NbSe$_2$ for superconducting hybrid devices, enabling both zero- and one-dimensional electrostatically-defined confined regions in close proximity to a superconductor.
Using gate-defined quantum dots as local spectroscopic probes, we directly resolve the proximity-induced superconducting gap and find that the suppression of the density of states persists over distances exceeding 1~$\mu$m.
These results establish BLG/NbSe$_2$ heterostructures as a versatile platform for combining electrostatic confinement with superconducting proximity effects. 
This opens a route toward more complex hybrid devices, including coupled quantum dots for Cooper-pair splitting and gate-tunable Andreev-state spectroscopy.

~\\
\textbf{Supporting Information.} Details on sample fabrication and measurement setup, extraction of the gate lever arm, superconducting characterization of the heterostructure, additional tunneling and Coulomb spectroscopy data, extraction of the proximity-induced superconducting gap, and analysis of finite-size quantization in the proximitized bilayer graphene lead.

\begin{acknowledgments}
The authors thank L. Banszerus and V. Mourik for fruitful discussions, F. Lentz and S. Trellenkamp for help with sample fabrication and E. Neumann for support with SEM-EDX analysis.
This project has received funding from the Deutsche Forschungsgemeinschaft (DFG, German Research Foundation) under Germany’s Excellence Strategy – Cluster of Excellence Matter and Light for Quantum Computing (ML4Q) EXC 2004/1 and EXC 2004/2 – 390534769, by the FLAG-ERA grant ThinQ, by the Deutsche Forschungsgemeinschaft (DFG, German Research Foundation) - 534269806, the Helmholtz Nano Facility~\cite{albrecht2017may}, and from OTKA K138433, FlagERA MultiSpin, COST Action CA 21144 superqumat.
K.W. and T.Ta. acknowledge support from the CREST (JPMJCR24A5), JST and World Premier International Research Center Initiative (WPI), MEXT, Japan. S.C. was supported by the Alexander von Humboldt Foundation.

\end{acknowledgments}
~\\
\textbf{Data availability} The data supporting the findings are available in a Zenodo repository under https://doi.org/10.5281/zenodo.22751091.

\end{document}


\title{
\bf Supporting Information: \\ 
Probing proximity-induced superconductivity in bilayer graphene using gate-defined quantum dots 
}

\author{Philipp Schmidt}
\email[]{philipp.schmidt3@rwth-aachen.de}
\affiliation{JARA-FIT and 2nd Institute of Physics, RWTH Aachen University, 52074 Aachen, Germany}
\affiliation{Peter Gr\"unberg Institute (PGI-9), Forschungszentrum J\"ulich, 52425 J\"ulich, Germany}
\author{Mathies Funke}
\affiliation{JARA-FIT and 2nd Institute of Physics, RWTH Aachen University, 52074 Aachen, Germany}
\author{Till Tenger}
\affiliation{JARA-FIT and 2nd Institute of Physics, RWTH Aachen University, 52074 Aachen, Germany}
\author{Katarina Stanojevi\'{c}}
\affiliation{JARA-FIT and 2nd Institute of Physics, RWTH Aachen University, 52074 Aachen, Germany}
\author{Hubert Dulisch}
\affiliation{JARA-FIT and 2nd Institute of Physics, RWTH Aachen University, 52074 Aachen, Germany}
\affiliation{Peter Gr\"unberg Institute (PGI-9), Forschungszentrum J\"ulich, 52425 J\"ulich, Germany}
\author{Kenji Watanabe}
\affiliation{Research Center for Electronic and Optical Materials, National Institute for Materials Science, 1-1 Namiki, Tsukuba 305-0044, Japan}
\author{Takashi Taniguchi}
\affiliation{ Research Center for Materials Nanoarchitectonics, National Institute for Materials Science,  1-1 Namiki, Tsukuba 305-0044, Japan}
\author{Fabian Hassler}
\affiliation{Institute for Quantum Information, RWTH Aachen University, 52056 Aachen, Germany}
\author{Christian Volk}
\affiliation{JARA-FIT and 2nd Institute of Physics, RWTH Aachen University, 52074 Aachen, Germany}
\affiliation{Peter Gr\"unberg Institute (PGI-9), Forschungszentrum J\"ulich, 52425 J\"ulich, Germany}
\author{Szabolcs Csonka}
\affiliation{Department of Physics, Institute of Physics, Budapest University of Technology and Economics, M\"uegyetem rkp. 3., H-1111 Budapest, Hungary}
\affiliation{MTA-BME Superconducting Nanoelectronics Momentum Research Group, M\"uegyetem rkp. 3., H-1111 Budapest, Hungary}
\author{Christoph Stampfer}
\affiliation{JARA-FIT and 2nd Institute of Physics, RWTH Aachen University, 52074 Aachen, Germany}
\affiliation{Peter Gr\"unberg Institute (PGI-9), Forschungszentrum J\"ulich, 52425 J\"ulich, Germany}

\maketitle


\section{Sample fabrication and measurement setup}

\begin{figure}[h]
    \centering
    \includegraphics[width=0.99\linewidth]{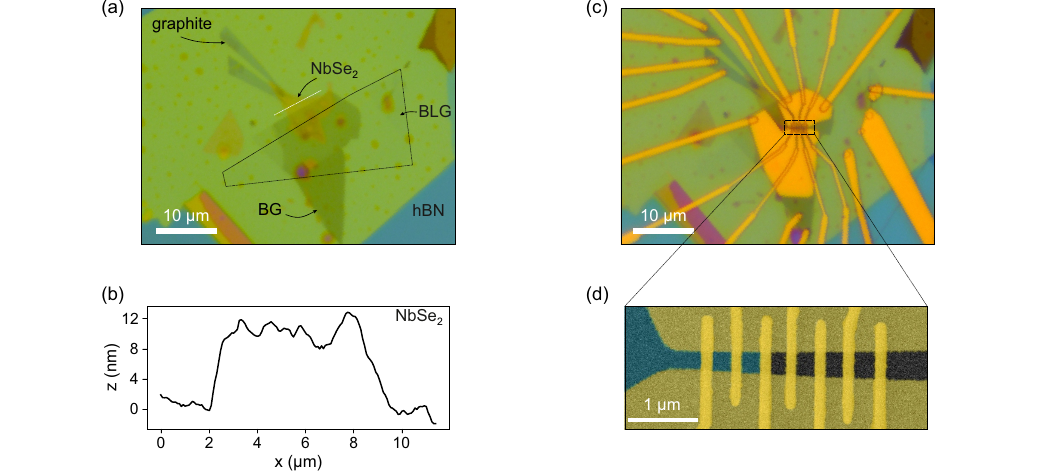}
    \caption{(a) Optical microscopy image of the 2D van der Waals heterostructure. Two graphite leads contact the NbSe$_2$ which is partially overlapping with the BLG on top of the graphite back gate (BG). 
    (b) Height profile of the NbSe$_2$ flake along the white dashed line in panel (a) measured by atomic force microscopy (AFM) indicating a flake thickness of $\approx 10\,\mathrm{nm}$.
    (c) Optical microscopy image of the finished device after fabrication of the ohmic contacts and deposition of all gate layers.
    (d) False colored scanning electron microscopy image of the channel area indicated in panel (c).
    }
    \label{fig:Sample}
\end{figure}

Thin flakes of hexagonal boron nitride (hBN), bilayer graphene (BLG), few layer graphite, and niobium diselenide (NbSe$_2$) were prepared via mechanical exfoliation.
The graphite crystals, from which BLG and few-layer graphite flakes were exfoliated, were commercially sourced from NGS Naturgraphit, while bulk NbSe$_2$ crystals were obtained from HQ Graphene. The hBN crystals were provided by T. Taniguchi and K. Watanabe.
These flakes were identified through optical contrast under a microscope, with verification provided by a machine learning method that accounts for thickness-dependent optical contrast and flake geometry \cite{uslu2024feb}.
We note that the NbSe$_2$ was exfoliated onto a HMDS coated Si/SiO$_2$ wafer helping in detaching the flake from the substrate later.
Furthermore, the graphite flake which will be used as an ohmic contact to the NbSe$_2$ flake is patterned by electron beam lithography (EBL) and consecutive reactive ion etching (RIE) into thin strips. 
This enables us to measure the actual voltage drop over the NbSe$_2$/BLG structure.
Next, using these flakes, the van der Waals heterostructure is assembled by means of a polymer-based dry transfer method \cite{wang2013nov}, and placed onto a Si$^{++}$ substrate covered by 285\,nm of SiO$_2$.
The final heterostructure consists of the graphite-contacted NbSe$_2$ flake which is partially overlapping with the BLG, hBN that encapsulates it, and a graphite flake at the bottom acting as a back gate, see Fig.~\ref{fig:Sample}(a).
The thickness of the NbSe$_2$ flake is $\approx 10\,\mathrm{nm}$, see height profile in Fig.~\ref{fig:Sample}(b), and the BLG/NbSe$_2$ overlap area is $\approx 20\,\mathrm{\mu m^2}$.
The graphite leads as well as the BLG are electrically contacted by RIE and evaporation of 5\,nm Cr and 70\,nm Au.
Finally, the split gates and finger gates, separated by an insulating layer of Al$_2$O$_3$, are fabricated by gold deposition of 50\,nm and 70\,nm, respectively.
An optical microscopy image and a scanning electron microscopy image of the final device are shown in Figs.~\ref{fig:Sample}(c) and~\ref{fig:Sample}(d).

\begin{figure}[h]
    \centering
    \includegraphics[width=0.99\linewidth]{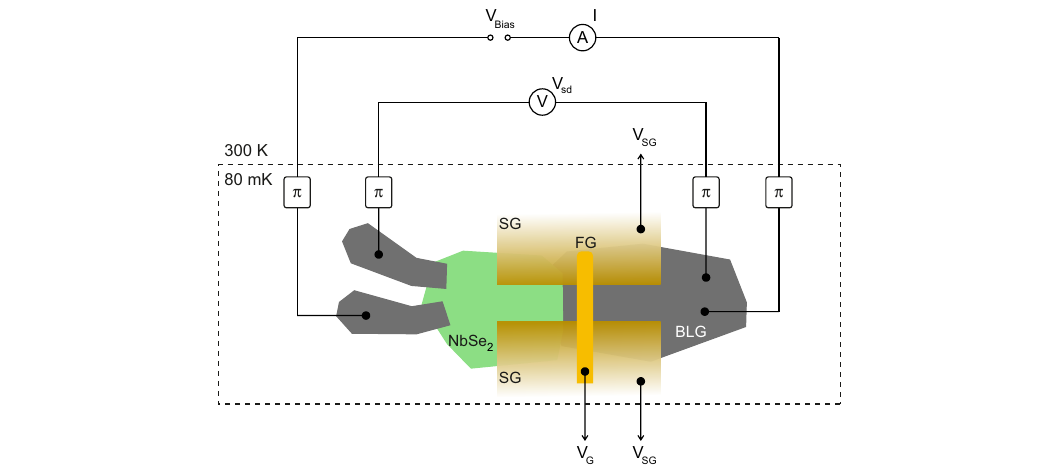}
    \caption{Simplified schematic drawing of the experimental setup. The device is measured in a four-terminal configuration using DC and low frequency lock-in techniques. All lines are filtered by QDevil $\pi$-filters at the cryostats mixing chamber plate.
    }
    \label{fig:Setup}
\end{figure}

Transport measurements were performed in a dilution refrigerator with a base temperature of 80\,mK, employing both DC and low-frequency lock-in detection methods. 
Current was measured using an I-V converter, while voltage measurements were carried out in a four-terminal configuration with a low-noise differential amplifier. 
All electrical lines were filtered at the mixing chamber stage of the cryostat using QDevil filters. 
A simplified schematic of the experimental setup is presented in Fig.~\ref{fig:Setup}.
We note, that a constant parasitic parallel conductance of $\approx0.005\,e^2/h$ for DC and $\approx0.1\,e^2/h$ for lock-in measurements is subtracted in data post processing. 
Selected datasets were lightly smoothed to reduce experimental noise and improve the visibility of spectroscopic features. Savitzky–Golay filtering was applied along the source–drain bias ($V_{\mathrm{sd}}$) direction. The filter parameters are specified as (window length in data points, polynomial order): Fig. 2(s): (5, 3); Fig. 3(e): (3, 2); Fig. 3(g): (19, 1); and Fig. 4(a): FG1 (3, 2), FG3 (3, 1), and FG4 (3, 1). A two-dimensional median filter with a 2 × 2 data-point kernel was applied to Figs. 3(b,c), while a two-dimensional Gaussian filter with a standard deviation of 1 data point along each axis was applied to Fig. 4(a), FG2. All other datasets are shown without smoothing.

\newpage
\section{Extraction of the gate lever arm}

\begin{figure}[h]
    \centering
    \includegraphics[width=0.99\linewidth]{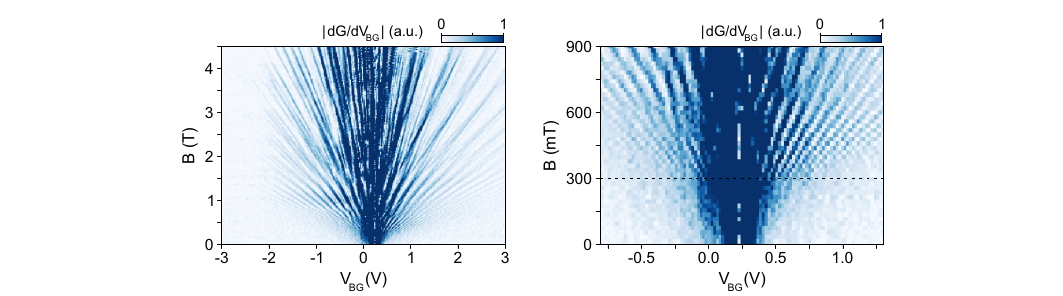}
    \caption{Derivative of the conductance, $|\mathrm{d}G/\mathrm{d}V_\mathrm{BG}|$, as a function of back gate voltage $V_\mathrm{BG}$ and magnetic field $B$, measured at $80\,\mathrm{mK}$. All split gate and finger gate voltages are set to $0\,\mathrm{V}$.}
    \label{fig:SdH}
\end{figure}

The gate-voltage-dependent charge carrier density in the electrostatically-defined channel is determined  by $n_\mathrm{ch} = \alpha_\mathrm{BG} V_\mathrm{BG}$ where $V_\mathrm{BG}$ is the back gate voltage and $\alpha_\mathrm{BG}$ is the respective gate lever arm.
We extract the gate lever arm from the slope of lines corresponding to a constant filling factor in the quantum Hall measurements (see Fig.~\ref{fig:SdH}), yielding $\alpha_\mathrm{BG} = 5.74 \times 10^{15}\,\mathrm{V^{-1}m^{-2}}$.

\section{Superconducting properties of the heterostructure}

\begin{figure}[h]
    \centering
    \includegraphics[width=0.99\linewidth]{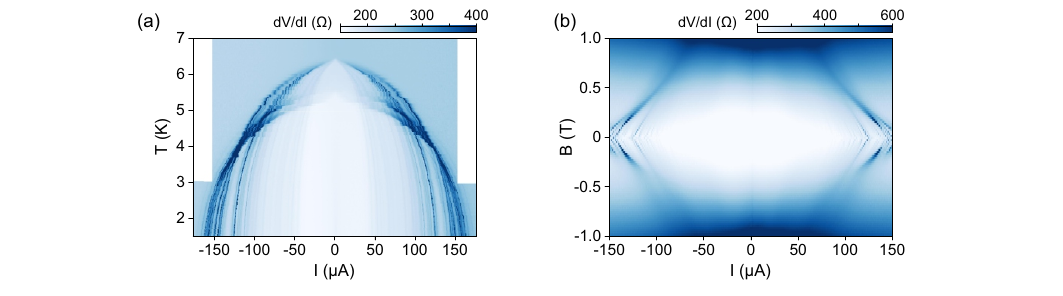}
    \caption{Differential resistance $\mathrm{d}V/\mathrm{d}I$ of the NbSe$_2$/BLG heterostructure as a function of current $I$ and (a) temperature $T$ at $B=0\,\mathrm{T}$ or (b) magnetic field $B$ at $T=80\,\mathrm{mK}$. All gate voltages are set to $0\,\mathrm{V}$.}
    \label{fig:vccs}
\end{figure}

We perform a basic characterization of the NbSe$_2$ / BLG heterostructure to investigate its superconducting properties.
Measurements of differential resistance as a function of bias current and temperature (Fig.~\ref{fig:vccs}(a)) reveal a superconducting dome with a critical temperature $T_c \approx 6.5$\,K, consistent with previously reported values for NbSe$_2$ flakes of similar thickness \cite{khestanova2018apr}.
Further analysis of the differential resistance under application of an out-of-plane magnetic field suggests a critical field exceeding 1\,T, see Fig.~\ref{fig:vccs}(b).
However, an accurate determination of the critical field is hindered by the dominant magnetoresistance response of the BLG, which masks the suppression of superconductivity.

\newpage
\section{Additional tunneling spectroscopy data}

\begin{figure}[h]
    \centering
    \includegraphics[width=0.99\linewidth]{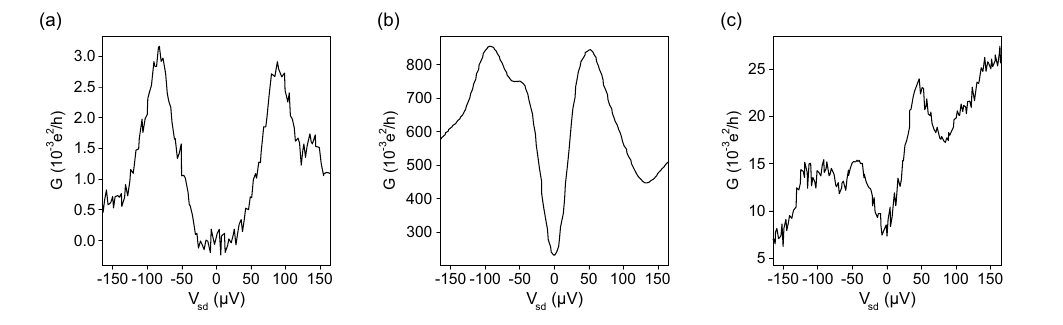}
    \caption{Differential conductance $G$ as a function of source-drain voltage $V_\mathrm{sd}$ in the tunneling regime, measured using (a)~FG~1 at $V_\mathrm{G1}=3.514\,\mathrm{V}$, (b)~FG~2 at $V_\mathrm{G2}=3.388\,\mathrm{V}$, and (c)~FG~3 at $V_\mathrm{G3}=3.676\,\mathrm{V}$. The back gate and split gate voltages are fixed at $V_\mathrm{BG}=-2.4\,\mathrm{V}$ and $V_\mathrm{SG}=2.9\,\mathrm{V}$, respectively, for all measurements.}

    \label{fig:tunneling_vs_fg}
\end{figure}

\begin{figure}[h]
    \centering
    \includegraphics[width=0.99\linewidth]{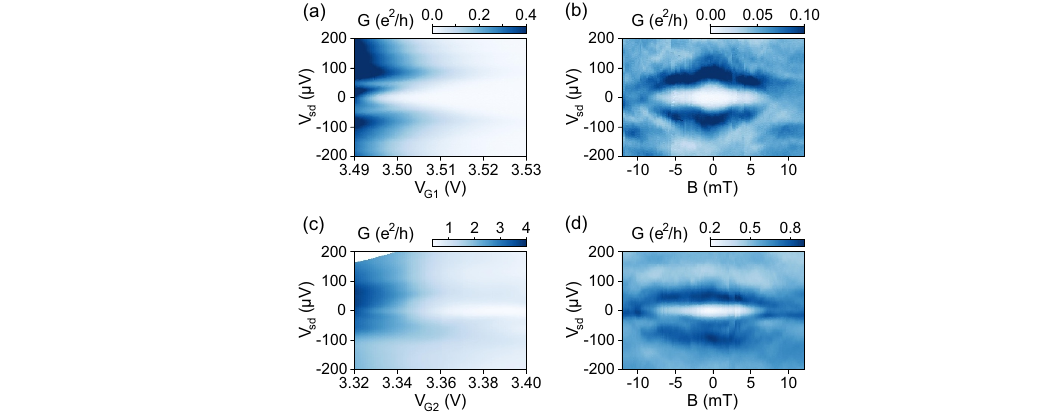}
    \caption{Differential conductance $G$ as a function of source-drain voltage $V_\mathrm{sd}$ and (a,c) finger gate voltage or (b,d) magnetic field in the tunneling regime, measured using FG\,1 (a,b) and FG\,2 (c,d). The back gate and split gate voltages are fixed at $V_\mathrm{BG}=-2.4\,\mathrm{V}$ and $V_\mathrm{SG}=2.9\,\mathrm{V}$, respectively, for all measurements.}
    \label{fig:tunneling_maps}
\end{figure}

In addition to the tunneling spectroscopy data presented in the main text, we show tunneling line traces for different FGs in Fig.~\ref{fig:tunneling_vs_fg}.
Furthermore, we present gate voltage and magnetic field dependencies of the tunneling spectroscopy, measured at a back gate voltage of $V_\mathrm{BG}=-2.4\,\mathrm{V}$ for FG~1 and 2 (Fig.~\ref{fig:tunneling_maps}).

\begin{figure}[h]
    \centering
    \includegraphics[width=0.99\linewidth]{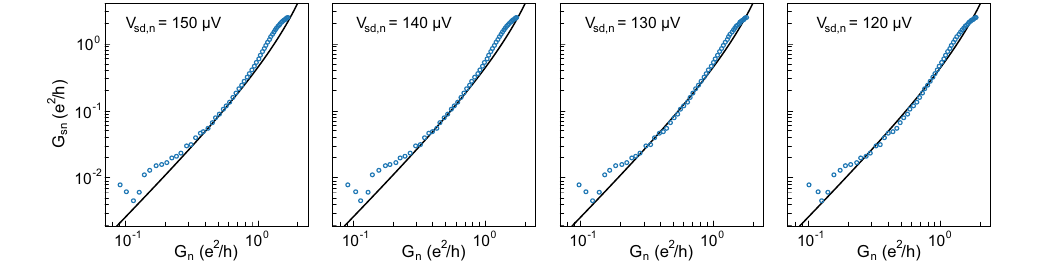}
    \caption{Evaluation of the data shown in Fig.~2(f) of the main text using different bias voltages between $120$ and $150\,\mu\mathrm{V}$ to define the normal-state conductance $G_n$.}
    \label{fig:Fig_S_loglog_multiBias}
\end{figure}

\begin{figure}[h]
    \centering
    \includegraphics[width=0.99\linewidth]{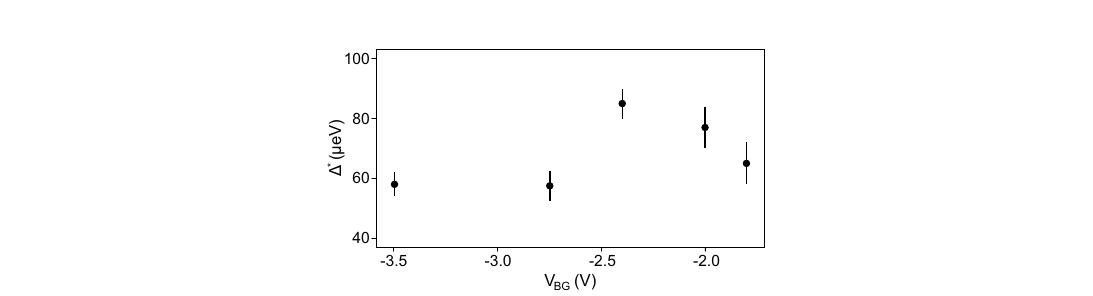}
    \caption{Proximity-induced superconducting gap $\Delta^*$ as a function of back gate voltage $V_\mathrm{BG}$, extracted from the tunneling measurements shown in Fig.~1(d) of the main text.}
    \label{fig:Gap_vs_VBG}
\end{figure}

\newpage
\section{Extraction of the induced superconducting gap}

The induced superconducting gap $\Delta^\ast$ is extracted from differential conductance traces measured as a function of the source--drain bias voltage $V_{\mathrm{sd}}$. The tunneling spectra exhibit two conductance peaks at positive and negative source--drain bias voltages, corresponding to the superconducting coherence peaks.
To determine the gap, we identify the source--drain bias voltages $V_{\mathrm{sd},1}$ and $V_{\mathrm{sd},2}$ at which the coherence peaks occur. The peak-to-peak separation is defined as $\Delta V_{\mathrm{sd}} = |V_{\mathrm{sd},2} - V_{\mathrm{sd},1}|$. The induced superconducting gap is then calculated as
$$
\Delta^\ast = \frac{e}{2}\Delta V_{\mathrm{sd}},
$$
where $e$ is the elementary charge.

\begin{figure}[h]
    \centering
    \includegraphics[width=0.99\linewidth]{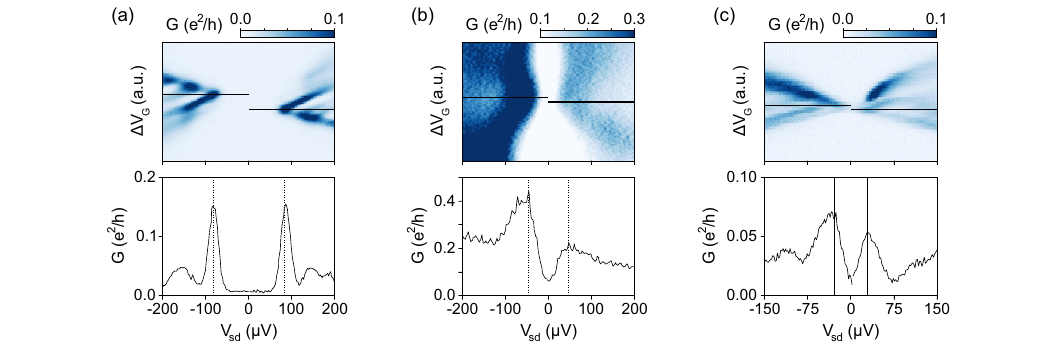}
    \caption{Extraction of the proximity-induced superconducting gap from Coulomb spectroscopy measurements of quantum dots defined by FGs 1 (a), 2 (b), and 3 (c). The superconducting gap is determined from the positions of the conductance peaks in the line traces shown in the lower panels, extracted along the solid black lines indicated in the corresponding upper panels.
    }
    \label{fig:Fig_S_QD_gap_extraction}
\end{figure}

For the quantum dots the induced superconducting gap is extracted from the displaced Coulomb diamond tips in the differential conductance maps, shown in Fig.~\ref{fig:Fig_S_QD_gap_extraction}.
Since the resonances are shifted in both source-drain bias and gate voltage, two line traces are taken for each dataset along the solid black lines indicated in the top panels, each following one branch of the displaced diamond tips. 
The proximity-induced gap $\Delta^*$ is obtained from the separation of the conductance peaks.

\newpage
\section{Additional Coulomb spectroscopy data}

\begin{figure}[h]
    \centering
    \includegraphics[width=0.99\linewidth]{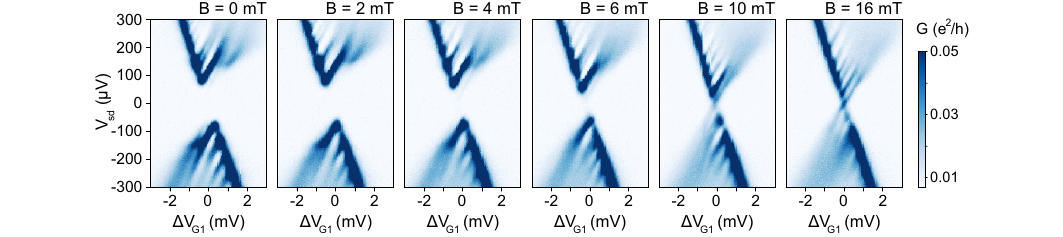}
    \caption{Extended sequence of the zoomed-in Coulomb spectroscopy measurements presented in Fig.~3 of the main text, showing additional magnetic field values.
    The measurements were performed using FG\,1, with the back gate and split gate voltages fixed at $V_\mathrm{BG}=-2.4\,\mathrm{V}$ and $V_\mathrm{SG}=2.9\,\mathrm{V}$, respectively.}
    \label{fig:qd_m2p4_bfield}
\end{figure}

\begin{figure}[h]
    \centering
    \includegraphics[width=0.99\linewidth]{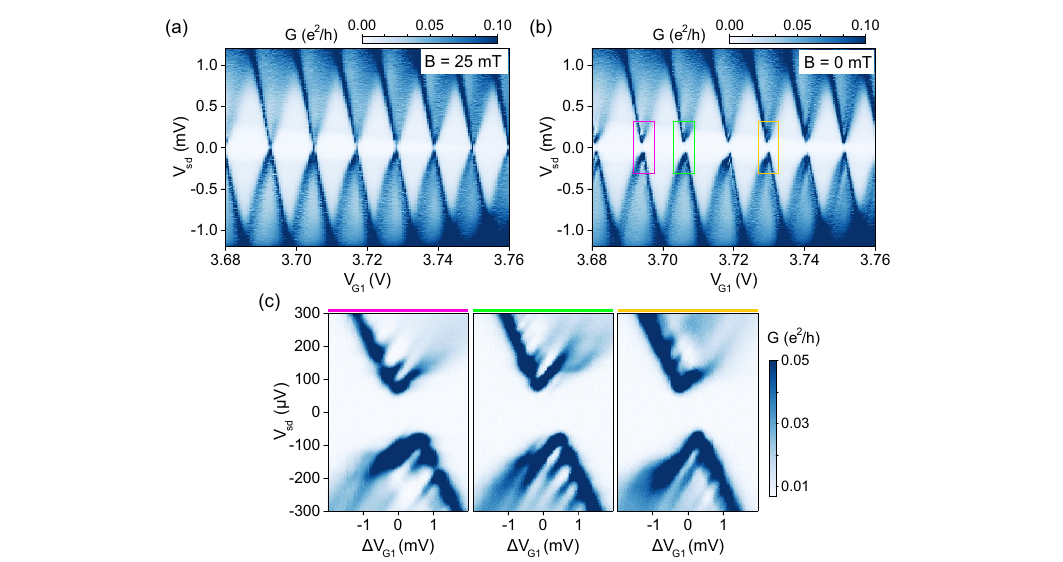}
    \caption{Additional dataset of Coulomb spectroscopy of a quantum dot formed with FG\,1, measured at a back gate voltage of $V_\mathrm{BG}=-2.4\,\mathrm{V}$ with applied magnetic field (a) and without magnetic field (b). (c) Zoom to different charge transitions marked by colored boxes in (b).
    }
    \label{fig:qd_m2p4}
\end{figure}

\begin{figure}[h]
    \centering
    \includegraphics[width=0.99\linewidth]{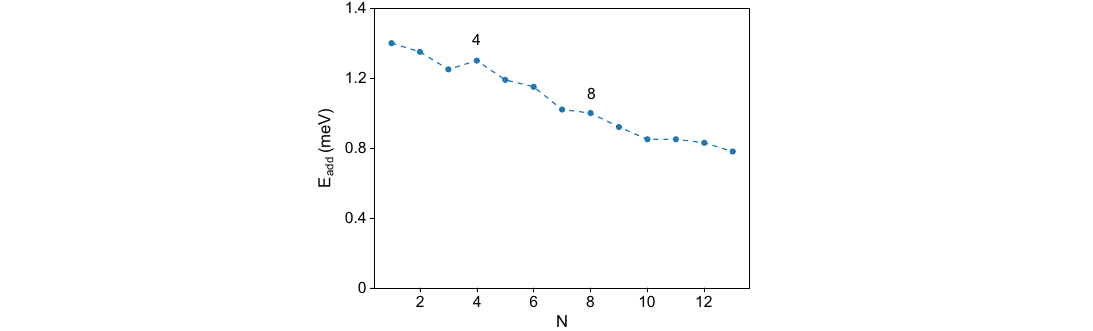}
    \caption{Quantum dot addition energy $E_\mathrm{add}$ as a function of the relative carrier number $N$, determined from the extends of the Coulomb diamonds shown in Fig.~3(b) of the main text. The measurements were performed at a magnetic field of $B = 25\,\mathrm{mT}$, where the device is in the normal state.
    }
    \label{fig:addition_energies}
\end{figure}

\begin{figure}[h]
    \centering
    \includegraphics[width=0.99\linewidth]{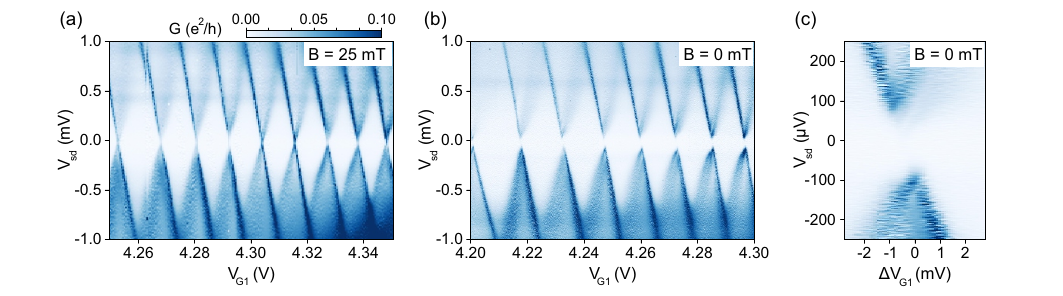}
    \caption{Differential conductance $G$ as a function of source-drain voltage $V_\mathrm{sd}$ and finger gate voltage $V_\mathrm{G1}$, showing Coulomb diamond spectroscopy of a quantum dot formed using FG\,1. The measurements were performed at a back gate voltage of $V_\mathrm{BG}=-2.75\,\mathrm{V}$ and a split gate voltage of $V_\mathrm{SG}=3.35\,\mathrm{V}$, with an applied magnetic field of (a) $B=25\,\mathrm{mT}$ and (b,c) $B=0\,\mathrm{mT}$.}
    \label{fig:qd_m2p75}
\end{figure}

\begin{figure}[h]
    \centering
    \includegraphics[width=0.99\linewidth]{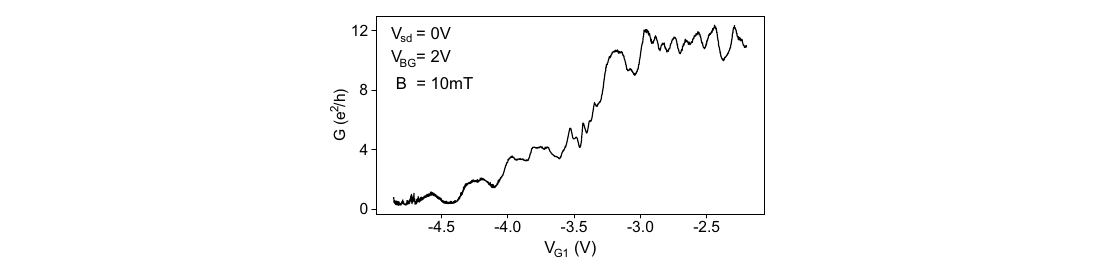}
    \caption{Differential conductance at  $V_\mathrm{BG} = 2\,\mathrm{V}$, i.e. inverted channel polarity. Differential conductance $G$ as a function of finger gate voltage $V_\mathrm{G1}$, measured at a fixed back gate voltage of $V_\mathrm{BG} = 2\,\mathrm{V}$, source–drain bias $V_\mathrm{sd} = 0\,\mathrm{V}$, and an applied magnetic field of $B = 10\,\mathrm{mT}$.
    }
    \label{fig:Positive_BG}
\end{figure}

In addition to the Coulomb spectroscopy data presented in the main text, we provide an extended sequence of the zoomed-in measurements shown in Fig.~3, covering additional magnetic-field values (Fig.~\ref{fig:qd_m2p4_bfield}). We further present additional Coulomb spectroscopy datasets for a quantum dot formed using FG\,1 at back gate voltages of $V_\mathrm{BG}=-2.4\,\mathrm{V}$ (Fig.~\ref{fig:qd_m2p4}) and $V_\mathrm{BG}=-2.75\,\mathrm{V}$ (Fig.~\ref{fig:qd_m2p75}).
The addition energies extracted from the Coulomb diamonds in Fig.~3(b) of the main text are shown in Fig.~\ref{fig:addition_energies}. 
Finally, we present differential conductance measurements at inverted channel polarity, obtained at a positive back-gate voltage of $V_\mathrm{BG}=2\,\mathrm{V}$ (Fig.~\ref{fig:Positive_BG}).

\newpage
\section{Ballistic and diffusive considerations for the spatial proximity effect}
To further examine the spatial dependence of the proximity-induced superconducting gap, we compare the ballistic dwell-time model with a phenomenological exponential decay.
\\
In the ballistic picture, the characteristic excitation energy is determined by the time of flight between the superconducting interface and the electrostatically defined tunneling barrier,
\begin{equation}
\Delta^*(x)=\frac{\pi\hbar v}{4x},
\end{equation}
where $v$ is the effective longitudinal velocity. 
Using $v$ as the only fit parameter gives $v\simeq5.2\times10^4$~m/s, a physically reasonable value for gapped BLG close to the band edge.
\\
As an alternative description, we consider an exponential dependence,
\begin{equation}
\Delta^*(x)=\Delta_0^*\exp(-x/\xi),
\end{equation}
which has previously been employed to characterize the spatial decay of proximity-induced superconducting gaps~\cite{natterer2016jan,ganguli2022mar}.
The fit yields $\Delta_0^*\approx110$~$\mu$eV and $\xi\approx740$~nm. 
We emphasize that this dependence is phenomenological and does not constitute a microscopic description of diffusive transport. 
In a finite diffusive proximity region, the characteristic energy scale is instead related to the Thouless energy, $E_\mathrm{Th}\sim\hbar D/x^2$, where $D$ is the diffusion constant. 
A quantitative diffusive calculation would additionally require information about the interface transparency and the electrostatic boundary conditions. 
Given the limited number of spatial data points and the two free parameters of the exponential fit, the present data cannot reliably distinguish between ballistic and diffusive descriptions.
\\
Independent information about scattering in the BLG channel is obtained from Shubnikov--de Haas oscillations. At a carrier density of approximately $n=4.0\times10^{11}$~cm$^{-2}$, the oscillations become visible around $B\approx0.3$~T (see Fig.~S3). Using the approximate onset condition $\omega_c\tau_q\sim1$, we estimate the quantum mean free path as
\begin{equation}
l_q=v_F\tau_q\approx\frac{\hbar k_F}{eB}\approx250~\mathrm{nm},
\end{equation}
where we assume a nearly circular Fermi surface with $k_F=\sqrt{\pi n}$. This estimate is approximate, since the onset of quantum oscillations also depends on thermal broadening and experimental resolution. Moreover, the quantum scattering lifetime includes small-angle scattering, which contributes less strongly to momentum relaxation. The transport mean free path may therefore be considerably larger than $l_q$. 
\\
We further compare the exponential decay length with the ballistic superconducting length scale,
\begin{equation}
\xi_\mathrm{bal}=\frac{\hbar v_F}{\Delta_0^*}.
\end{equation}
\\
Using $v_F=5.2\times10^4$~m/s and $\Delta_0^*=110$~$\mu$eV gives $\xi_\mathrm{bal}\approx310$~nm, approximately a factor of $2.4$ smaller than the fitted exponential decay length of $\xi\approx740$~nm. This difference may partly reflect the distinction between the effective longitudinal velocity extracted from the dwell-time model and the Fermi velocity of the BLG channel. In particular, the effective longitudinal velocity depends on the quasiparticle trajectories and boundary conditions. A Fermi velocity of approximately $1.24\times10^5$~m/s would yield a ballistic length scale of $740$~nm. Without an independent determination of the relevant velocity, however, this comparison cannot establish the transport regime.
\\
Overall, the one-parameter ballistic description, the independently estimated quantum mean free path, and the observation of finite-size resonances favor a ballistic or quasi-ballistic interpretation. Nevertheless, additional scattering and a crossover toward diffusive transport, particularly at larger distances from the superconducting interface, cannot be excluded.

\newpage
\section{Finite-Size Quantization in the Lead Segment}

\begin{figure}[h]
    \centering
    \includegraphics[width=0.3\linewidth]{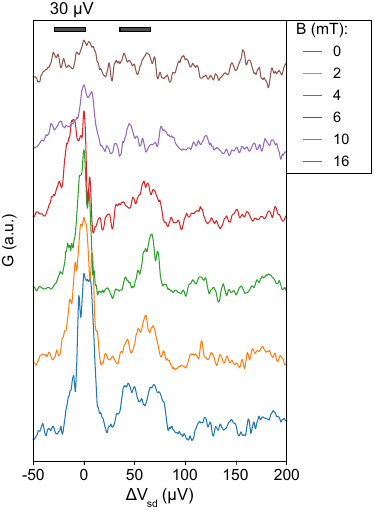}
    \caption{Conductance oscillations as function of voltage difference $\Delta V_\mathrm{sd}$ (with respect to the first oscillation maximum) for different magnetic fields extracted from the Coulomb spectroscopy presented in  Fig.~3 of the main text. The traces are vertically shifted for clarity.
    }
    \label{fig:oscillations}
\end{figure}

Conductance oscillations in the Coulomb diamond spectroscopy (cf. Fig.~3 of the main text) are shown as line traces for different magnetic fields, see Fig.~\ref{fig:oscillations}.

\begin{figure}[h]
    \centering
    \includegraphics[width=0.99\linewidth]{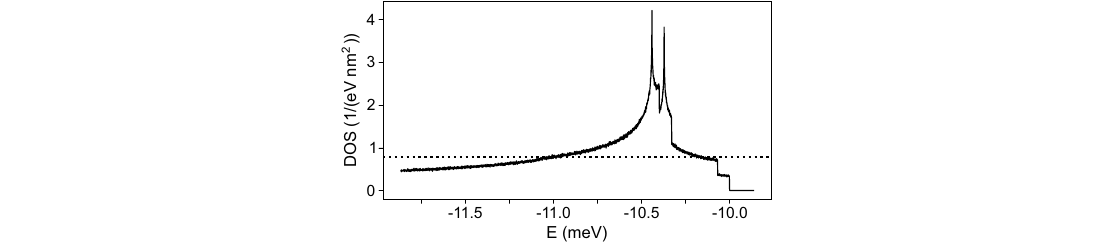}
    \caption{Density of states (DOS) of gapped bilayer graphene ($E_\mathrm{g}=20\,\mathrm{meV}$) as function of energy $E$ relative to the center of the band gap. The dashed line indicates the DOS value used for the estimate of the effective confinement length.
    }
    \label{fig:dos}
\end{figure}

To assess whether the finite-bias resonances discussed in the main text are consistent with finite-size quantization of the proximitized bilayer-graphene (BLG) lead segment, we estimate the characteristic confinement area from the calculated density of states (DOS) and the experimentally observed level spacing. From the applied gate voltages, we estimate a displacement field of $D/\epsilon_0 \approx 0.33~\mathrm{V/nm}$ in the relevant BLG region, corresponding to an induced band gap of approximately $E_{\mathrm{g}}\approx 20~\mathrm{meV}$. 
Using this gap, we calculate the BLG band structure based on the 8-band model parameterized by Konschuh et al.~\cite{konschuh2012mar} and obtain the corresponding DOS by Monte Carlo numerical integration over momentum space, as shown in Fig.~\ref{fig:dos}. In the energy range relevant to the experiment, the calculated density of states is approximately DOS$\approx 0.8~(\mathrm{eV\,nm^2})^{-1}$.

For a confined region of area $A$, the spacing between successive states can be estimated as $\delta E \simeq g/[DOS\times A]$, where $g=4$ accounts for the spin and valley degeneracy and $DOS$ includes both degeneracies. 
Using the experimentally extracted resonance spacing $\delta E\approx 55~\mu\mathrm{eV}$ yields $A\approx 9.1\times10^4~\mathrm{nm^2}$, corresponding to a characteristic linear dimension $\sqrt{A}\approx 300~\mathrm{nm}$. 
This length scale is in good agreement with the lithographic extent of the BLG lead segment between the quantum dot and the NbSe$_2$-covered region, supporting the interpretation of the observed resonances as a discrete spectrum arising from finite-size quantization in the proximitized lead. 
Given the uncertainty in the local displacement field and carrier energy, as well as the nontrivial geometry and confinement potential of the lead segment, this estimate is intended as an order-of-magnitude consistency check rather than a precise determination of the confinement dimensions.

\subsection{On the valley g-factor}
As shown in Fig. S15, additional resonances emerge at finite magnetic field. At $B = 16$ mT, we estimate a resonance splitting of approximately $30~ \mu eV$ (see black bars in Fig.~S15), corresponding to a valley g-factor of $g_v \approx 30$, using $\Delta E = g_v \mu_B B$. This value is consistent with previously reported valley g-factors of $20-40$ for comparatively large confinement areas (see Fig. 5(b) in Ref.~\cite{moller2023sep}), supporting the interpretation of these resonances as arising from valley splitting.